\documentclass[prl,twocolumn,superscriptaddress,nofootinbib]{revtex4-1}
\usepackage{epsfig}
\usepackage{graphicx}
\usepackage{palatino}
\usepackage[english]{babel}
\usepackage{hyphenat}
\usepackage{amsmath}
\usepackage{amssymb}
\usepackage{slashed}
\usepackage{epstopdf}
\usepackage{xcolor}
\usepackage{booktabs}
\usepackage[toc,page]{appendix}
\definecolor{lcolor}{rgb}{0.,0.0,0.}
\definecolor{citcolor}{rgb}{0,0.,0.5}
\usepackage[breaklinks,colorlinks,urlcolor=blue,citecolor=blue,linkcolor=blue]{hyperref}
\usepackage{multirow}
\usepackage{ltablex}

\newcommand{\beq}{\begin{equation}}
\newcommand{\eeq}{\end{equation}}
\newcommand{\bea}{\begin{eqnarray}}
\newcommand{\eea}{\end{eqnarray}}

\newcommand{\bem}{\begin{multline}}
\newcommand{\eem}{\end{multline}}
\newcommand{\beg}{\begin{gather}}
\newcommand{\eeg}{\end{gather}}

\newcommand{\ben}{\begin{eqnarray*}}
\newcommand{\een}{\end{eqnarray*}}

\newcommand{\eq}[1]{\begin{align}#1\end{align}}
\newcommand{\bal}{\begin{align}}
\newcommand{\eal}{\begin{align}}

\def\and{\quad\text{and}\quad}

\def\0{{\boldsymbol 0}}

\newcommand{\wt}{\widetilde}

\begin{document}
\title{{\bf Influence of the neutron-skin effect on nuclear isobar collisions at RHIC\\}}
\author{Jan Hammelmann}
\email[]{hammelmann@fias.uni-frankfurt.de}
\affiliation{Goethe-Universit\"at Frankfurt, Max-von-Laue-Strasse 1, 60438 Frankfurt am Main, Germany.}
\affiliation{Frankfurt Institute for Advanced Studies, Ruth-Moufang-Strasse 1, 60438 Frankfurt am Main, Germany.}
\author{Alba Soto-Ontoso}
\email[]{ontoso@bnl.gov}
\affiliation{Physics Department, Brookhaven National Laboratory, Upton, NY 11973, USA.}
\author{Massimiliano Alvioli}
\email[]{massimiliano.alvioli@irpi.cnr.it}
\affiliation{Consiglio Nazionale delle Ricerche, Istituto di Ricerca per la Protezione Idrogeologica, via Madonna Alta 126, I-06128 Perugia, Italy.}
\author{Hannah Elfner}
\email[]{elfner@fias.uni-frankfurt.de }
\affiliation{GSI Helmholtzzentrum f\"ur Schwerionenforschung GmbH, Planckstr. 1, 64291 Darmstadt, Germany.}
\affiliation{Goethe-Universit\"at Frankfurt, Max-von-Laue-Strasse 1, 60438 Frankfurt am Main, Germany.}
\affiliation{Frankfurt Institute for Advanced Studies, Ruth-Moufang-Strasse 1, 60438 Frankfurt am Main, Germany.}
\author{Mark Strikman}
\email[]{mxs43@psu.edu}
\affiliation{Pennsylvania State University, University Park, PA 16802, USA.}

\begin{abstract}
 The unambiguous observation of a Chiral Magnetic Effect (CME)-driven charge separation is the core aim of the isobar program at RHIC consisting of ${^{96}_{40}}$Zr+${^{96}_{40}}$Zr and ${^{96}_{44}}$Ru+${^{96}_{44}}$Ru collisions at $\sqrt {s_{\rm NN}}\!=\!200$ GeV. We quantify the role of the spatial distributions of the nucleons in the isobars on both eccentricity and magnetic field strength within a relativistic hadronic transport approach (SMASH, Simulating Many Accelerated Strongly-interacting Hadrons). In particular, we introduce isospin-dependent nucleon-nucleon spatial correlations in the geometric description of both nuclei, deformation for ${^{96}_{44}}$Ru and the so-called neutron skin effect for the neutron-rich isobar \textit{\textit{i.e.}} ${^{96}_{40}}$Zr. The main result of this study is a reduction of the magnetic field strength difference between ${^{96}_{44}}$Ru+${^{96}_{44}}$Ru and ${^{96}_{40}}$Zr+${^{96}_{40}}$Zr by a factor of 2, from $10\%$ to $5\%$ in peripheral collisions when the neutron-skin effect is included. Further, we find an increase of the eccentricity ratio between the isobars by up to 10\% in ultra-central collisions as due to the deformation of ${^{96}_{44}}$Ru while neither the neutron skin effect nor the nucleon-nucleon correlations result into a significant modification of this observable with respect to the traditional Woods-Saxon modeling. Our results suggest a significantly smaller CME signal to background ratio for the experimental charge separation measurement in peripheral collisions with the isobar systems than previously expected.
 
\end{abstract}

\maketitle
\section{Introduction}
One of the fundamental properties of Quantum Chromodynamics (QCD) is the axial anomaly, which in the massless fermion limit, reads as follows
\eq{\partial_\mu j^{\mu}_5=-\displaystyle\frac{g^2}{16\pi^2}F^a_{\mu\nu}\wt{F}^{a,{\mu\nu}}
\label{Eq1}}
where $F^a_{\mu\nu}$ is the gluon field strength, $\wt{F}^{a,{\mu\nu}}$ its dual, $g$ the strong coupling constant and $j_5^\mu$ the axial current density. The axial anomaly establishes a direct relationship between the generation of a net axial charge and the dynamics of non-Abelian gauge fields. 

Together with condensed matter systems~\cite{Li:2014bha}, ultra-relativistic heavy-ion collisions provide a unique environment to experimentally test the chiral anomaly. At least two mechanisms contribute to the right-hand side of Eq.~\ref{Eq1}. On the one hand, in the Color Glass Condensate description of the early, non-equilibrium stage of the collision known as Glasma~\cite{Lappi:2006fp}, fluctuations of the chromo-electric and chromo-magnetic fields give rise to a non-vanishing $F^a_{\mu\nu}\wt{F}^{a,{\mu\nu}}$~\cite{Lappi:2017skr,Guerrero-Rodriguez:2019ids}. Further, the non-trivial topological structure of the QCD vacuum results into another source of net axial charge density known as sphaleron transitions, whose rate is enhanced at high temperatures such as the ones reached in the Quark-Gluon Plasma phase~\cite{Klinkhamer:1984di, Mace:2016svc, Mace:2016shq}. These local fluctuations of axial charge density in the transverse plane occur in the presence of a strong electromagnetic field in non-central collisions~\cite{Stewart:2017zsu,Inghirami:2019mkc}. Then, the chiral imbalance is efficiently converted into a separation of positive and negative charges along the direction of the magnetic field. This phenomenon, dubbed Chiral Magnetic Effect (CME)~\cite{Fukushima:2008xe,Bzdak:2012ia}, manifests itself into charge-dependent azimuthal correlations of the measured hadrons~\cite{Voloshin:2004vk}.

A decade after the pioneering analysis of the STAR Collaboration~\cite{Abelev:2009ac}, the experimental confirmation of the CME remains unsettled. Numerous charge separation measurements in line with CME expectations were reported with different collisions systems and energies both from RHIC~\cite{STAR:2019xzd,Abelev:2009ad,Adamczyk:2014mzf} and LHC~\cite{Khachatryan:2016got,Abelev:2012pa}. However, these measurements are known to be strongly affected by background contamination arising from flow~\cite{Bzdak:2010fd} and local charge conservation~\cite{Schlichting:2010qia}. New observables beyond the traditional three particle correlator could help solving the problem~\cite{Magdy:2017yje}. Moreover, new RHIC measurements with different isobars, \textit{i.e.} ${^{96}_{40}}$Zr and ${^{96}_{44}}$R~\cite{Skokov:2016yrj} could disentangle the background from the signal. For that purpose, ${^{96}_{40}}$Zr+${^{96}_{40}}$Zr collisions will provide a precise characterization of the background contribution to the experimental charge separation measurement. On the other hand, the proton-rich isobar system ${^{96}_{44}}$Ru+${^{96}_{44}}$Ru will provide an enhanced sensitivity to the CME component, due to the formation of larger magnetic fields.

A correct interpretation of the forthcoming experimental data requires accurate quantification of background and signal from theory. A multiphase transport model predicted
the magnetic field strength, proportional to the CME contribution, to be $10\%$ larger for ${^{96}_{44}}$Ru+${^{96}_{44}}$Ru than for ${^{96}_{40}}$Zr+${^{96}_{40}}$Zr in peripheral collisions~\cite{Deng:2018dut,Zhao:2019crj}. A hydrodynamic framework predicted differences of up to $10\%$ on the elliptic flow of both collision systems related to deformation~\cite{Schenke:2019ruo}. A systematic comparison between a Woods-Saxon shape and density functional theory calculations shows that the functional form of the nuclear density distributions used in the simulations also impacts $v_2$ ( $\!\sim\!3\%$)~\cite{Xu:2017zcn}. All in all, the results of these studies identify the nuclear structure of the two isobar nuclei to be a source of uncertainty for $v_2$ but not for the magnetic field strength. 

In this work, we analyze the effect due to an experimentally measured nuclear phenomenon in the description of the density distribution of ${^{96}_{40}}$Zr \textit{i.e.} the neutron-skin effect~\cite{Trzcinska:2001sy, Trzcinska:2004dx}. This ingredient leads to an enhancement of the magnetic field in peripheral ${^{96}_{40}}$Zr+${^{96}_{40}}$Zr collisions within SMASH~\cite{Weil:2016zrk} consequently undermining the experimental prospects of finding out the Chiral Magnetic Effect with the isobar run.

\section{Framework}
\subsection{Neutron-skin effect and nucleon-nucleon correlations}
Traditionally, the spatial distribution of nucleons inside nuclei is generated by randomly sampling the Woods-Saxon density distribution~\cite{Woods:1954zz}
\beq
\rho(r,\theta)=\displaystyle\frac{\rho_0}{e^{(r-R'(\theta,\phi))/d}+1}
\label{WS1}
\eeq
where
\beq
R'(\theta) = R_0(1+\beta_2 Y_{2}^{0}(\theta)).
\label{WS2}
\eeq
In Eqs.~\ref{WS1}-\ref{WS2}, $\rho_0\!=\!0.168$ is the ground state density, $d$ refers to the difussiveness, $R_0$ is the nuclear radius and $\beta_2$ together with the spherical harmonic $Y_{2}^{0}$ control the deformation. Two severe simplifications are commonly made when using Eq.~\ref{WS1} for the nuclear geometry. First, nucleons are considered to be independent of each other. Second, protons and neutrons are treated indistinctly so that they are sampled from the same Woods-Saxon distribution \textit{i.e.} with identical values for $R_0$ and $d$. Experimental measurements and theoretical calculations ruled out both assumptios, as discussed below.

Since the early 80's the tails of the proton ($p$) and neutron ($n$) distributions are known to be distinct~\cite{Chaumeaux:1978pm,Horowitz:2013wha,Hagen:2015yea} \textit{i.e.} $R_0$ and $d$ in Eq.~\ref{WS1} are isospin dependent. The neutron distribution populates the outer region of neutron-rich nuclei. That is, the difference between the neutron and proton distributions mean square radii, which can be written as follows:
\beq
\Delta r_{np} = \langle r_n ^2\rangle^{1/2}-\langle r_p^2 \rangle^{1/2},
\eeq
is positive. Following the Woods-Saxon parametrization given by Eq.~\ref{WS1}, this phenomenon translates into nuclei having either $R_{0,p}\!>\!R_{0,n},d_{n}\!\sim\!d_{p}$ dubbed {\it{neutron-skin}} type or $R_{0,p}\!\sim\!R_{0,n},d_{n}\!>\!d_{p}$ refer to as {\it{neutron-halo}} type. A remarkable example of the latter category is $^{208}$Pb with $\Delta r_{np}\sim 0.15$~fm~\cite{Tarbert:2013jze}. In this case, the implications of $\Delta r_{np}\!\neq\!0$ in the context of observables relevant for the heavy-ion program at the LHC were recently studied in~\cite{Helenius:2016dsk,Paukkunen:2015bwa,Loizides:2017ack,Alvioli:2018jls}. Interestingly, one of the nuclei chosen for the isobar run at RHIC, $^{96}_{40}$Zr, also pertains to the neutron-halo category with
\beq
\Delta r_{np}\Big\vert_{^{96}_{40}{\rm Zr}} = 0.12\pm0.03~{\rm {fm}}
\label{deltarnp_zr}
\eeq
as extracted from the experimental analysis performed with the Low Energy Antiproton Ring at CERN~\cite{Trzcinska:2001sy,Trzcinska:2004dx}. The goal of this work is to study the consequences of considering isospin-dependent Woods-Saxon distributions fulfilling the upper limit of the constraint given by Eq.~\ref{deltarnp_zr}, $\Delta r_{np}\!=\!0.15$~fm, to describe $^{96}_{40}$Zr on CME-related observables. Note that we take the upper limit of $\Delta r_{np}\!=\!0.15$~fm with the purpose of studying the neutron-skin impact at its extreme. 

\begin{figure}[ht]
\begin{center}
\includegraphics[scale=0.55]{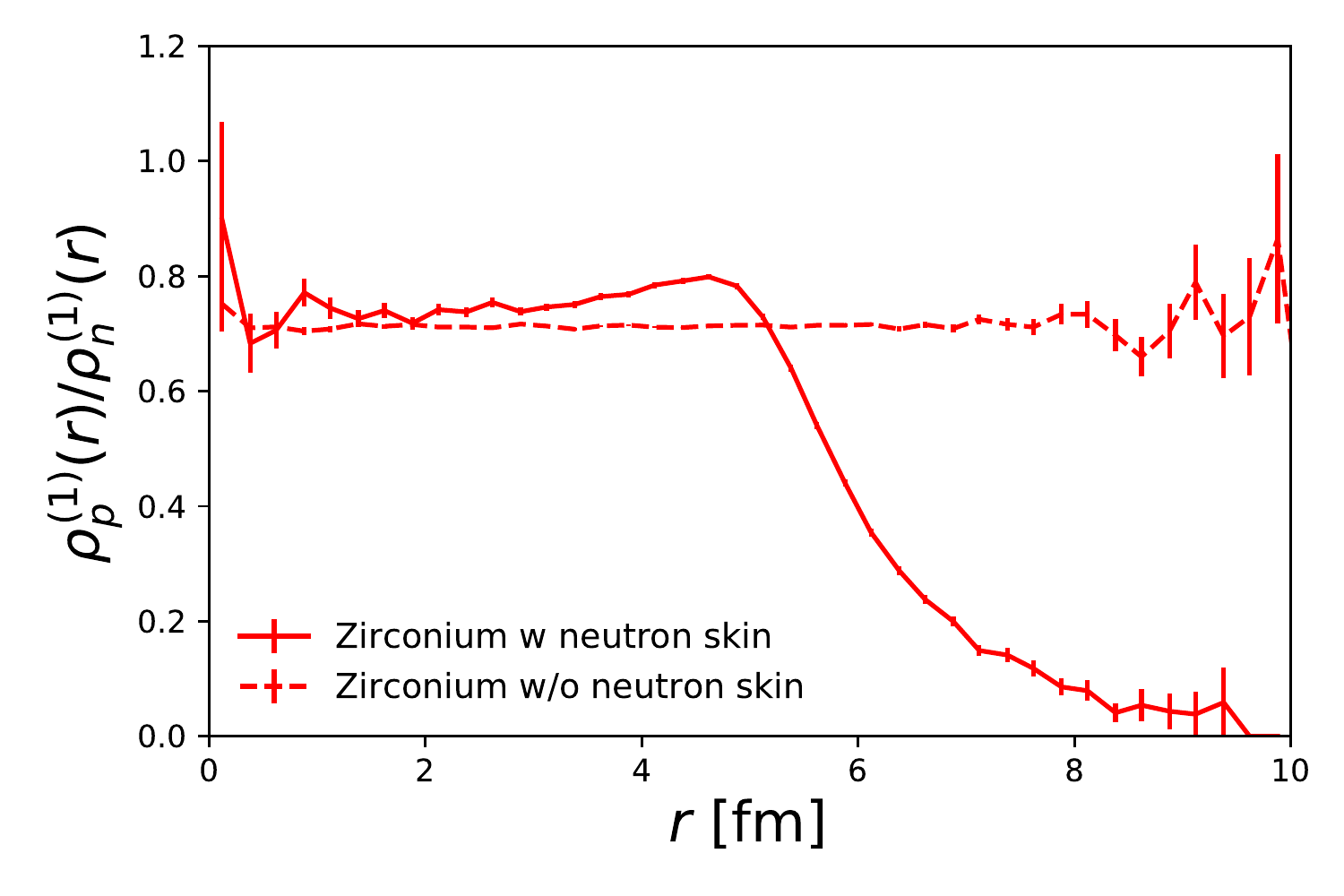} 
\end{center}
\vspace*{-0.5cm}
\caption[a]{Ratio of one-body density of protons to neutrons as a function of the radial distance for  $^{96}_{40}$Zr with (solid) or without (dashed) considering the neutron skin effect. Error bars account for statistical uncertainties.}
\label{1bratio}
\end{figure}

For that purpose, the starting point are the experimental values for ($R_0,d$) of the charge distribution~\cite{Pritychenko:2013gwa}, displayed both for $^{96}_{40}$Zr and $^{96}_{44}$Ru in Table~\ref{ChargeValues}. To extract the values of point distributions ($R_{0,p(n)},d_{p(n)}$) from the charged ones, keeping the size of the nucleus fixed, we follow the procedure outlined in~\cite{Patterson:2003wbr, Loizides:2017ack,Warda:2010qa}. It is not the goal of this paper to repeat the precise derivation detailed in the aforementioned papers. For completeness, in Appendix A we provide the formulae that were used to obtain the values of ($R_{0,p(n)},d_{p(n)}$) shown in Table~\ref{ZrValues} where we observe how $\Delta r_{np}\!=\!0.15$~fm is translated into a larger value of the diffusiveness for the neutron distribution while the radius remain the same for both types of nucleons. 
The ratio between the one body densities of neutrons and protons in $^{96}_{40}$Zr as function of the radial distance is displayed in Fig.~\ref{1bratio}.
As expected, this ratio remains flat for the Woods-Saxon distribution while the inclusion of the neutron-skin enhances the probability of finding a neutron inside the nucleus at large radial distances. Therefore, peripheral $^{96}_{40}$Zr+$^{96}_{40}$Zr collisions are expected to be dominated by neutron-neutron interactions. One further comment is in order before addressing the role of nucleon-nucleon (NN) short-range correlations (SRC). The nuclear structure of $^{96}_{40}$Zr and $^{96}_{44}$Ru differ not only because of the neutron-skin but also due to the deformation of the latter as exposed in Table~\ref{ChargeValues}. Although there are some studies (\textit{e.g.}~Ref.~\cite{Moller:2015fba}) that indicate the opposite situation \textit{i.e.} $^{96}_{40}$Zr is deformed while $^{96}_{44}$Ru is not, we stick to the former scenario in order to isolate the impact of deformation from the neutron-skin. The deformation affects the geometry of the nucleus in such a way that it has an ellipsoidal shape. In each event the deformed nuclei are rotated by an arbitrary angle before the collision to reflect the experimental situation in a realistic fashion.

\begin{table}[ht]
\centering
\begin{tabular}{|c||c|c|c|c}
\hline
Nucleus &$R_0$ [fm]&$d$ [fm] & $\beta_2$ \\
\hline\hline
$^{96}_{40}$Zr & 5.02 & 0.46 & 0 \\

$^{96}_{44}$Ru & 5.085 & 0.46 & 0.158 \\
\hline
\end{tabular}
\caption{Woods-Saxon parameters for the two isobar collision systems.}\label{ChargeValues}
\end{table}

\begin{table}[ht]
\centering
\begin{tabular}{|c||c|c|c}
\hline
Nucleon in $^{96}_{40}$Zr &$R_0$ [fm]&$d$ [fm]\\
\hline\hline
$p$ & 5.08 & 0.34\\

$n$ & 5.08 & 0.46\\
\hline
\end{tabular}
\caption{Woods-Saxon parameters for the proton and neutron distributions of $^{96}_{40}$Zr.}\label{ZrValues}
\end{table}

An accurate description of the colliding nuclei calls for, along with neutron skin, inclusion of NN correlations in the ground state. A few dedicated experiments unambiguously measured long-sought SRC in the last few years; see \textit{e.g.}~Ref.~\cite{Duer:2018sxh} and references therein. We expect SRC to play role in different nuclear phenomena, including high-energy processes involving nuclei~\cite{Alvioli:2008rw,Alvioli:2009iw}, the EMC effect due to the relative proton-neutron modifications of nuclear structure, asymmetric cold dense nuclear systems up to neutron stars.

Signatures of SRC correlations are a peculiar short-range structure, in coordinate space, and a proton-neutron pairs dominance with respect to like-nucleons pairs at high momenta, in momentum space~\cite{Alvioli:2013qyz}. In the context of the present study, we are interested in the spatial structure of the nuclear wave function. A full ab-initio theoretical description of the nuclear many-body wave function, especially for large nuclei, is an outstanding challenge.

To account for NN SRC in complex nuclei Alvioli et al.~\cite{Alvioli:2009ab} proposed a Metropolis Monte Carlo generator of nuclear configurations. The method is based on the use of an approximate wave function, including spatial and spin-isospin dependent correlation functions, as a prob- ability measure of the positions. A series of papers used configurations produced with this method for very different purposes. The most recent update is to account for neutron skin, provided a parametrization of the neutron and proton profiles is known~\cite{Alvioli:2018jls}, as in here.

\subsection{SMASH}
To demonstrate the effects of the deformation of $^{96}_{44}$Ru and the neutron skin of $^{96}_{40}$Zr in nuclear collisions, the hadronic transport approach SMASH is employed. As a reference to the calculations employing the sophisticated spatial distributions explained in the previous section the default Woods-Saxon initialisation as described in~\cite{Weil:2016zrk} is used. In SMASH all well established particles from the PDG 2018~\cite{Tanabashi:2018oca} data are included.  
Apart from the initialisation isospin symmetry is assumed, meaning that the masses of isospin partners are assumed to be equal as well as their interactions are identical. The collision criterion is realized in a geometric way. The initial binary interactions of nucleons at high $\sqrt{s}$ proceed mainly via string excitation and decay~\cite{Mohs:2019}. For all calculations SMASH-1.6 has been used~\cite{dmytro_oliinychenko_2019_3485108}. 

\section{CME-searches related observables}
\subsection{Background: Eccentricity}
\begin{figure}[ht]
\begin{center}
\includegraphics[scale=0.6]{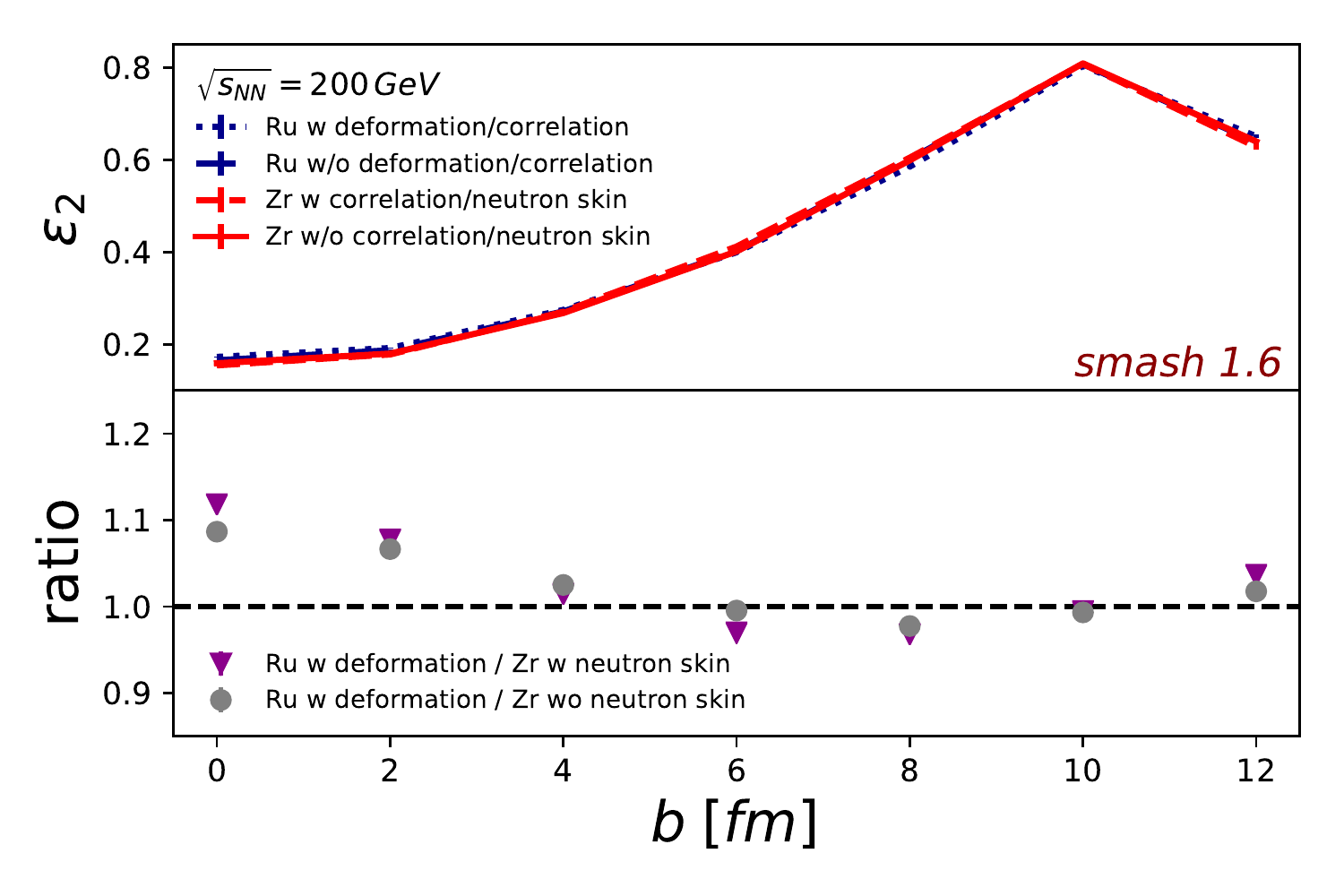} 
\end{center}
\vspace*{-0.5cm}
\caption[a]{Top: Participant eccentricity (see Eq.~\ref{epsilon2}) as a function of the impact parameter for $^{96}_{40}$Zr (red) with (solid) and without (dashed) neutron-skin and for $^{96}_{44}$Ru (blue) with (solid) and without (dashed) deformation. Bottom: Effect of the neutron skin on the eccentricity ratio between $^{96}_{44}$Ru and $^{96}_{40}$Zr.}
\label{eps2}
\end{figure}

The experimentally measured flow harmonics characterizing the azimuthal distribution of hadrons are an imprint of the QGP evolution acting on the initial spatial anisotropy of the nuclear overlap region. The latter is commonly characterized by the participant eccentricity defined, on an event-by-event basis, by
\beq
\varepsilon_2 = \frac{\sqrt{(\sigma_y-\sigma_x)^2+4\sigma_{xy}^2}}{\sigma_x^2 + \sigma_y^2}
\label{epsilon2}
\eeq
where $\sigma_x^2\!=\!\langle x^2 \rangle\!-\!\langle x \rangle^2$ and $\sigma_{xy}\!=\!\langle xy \rangle\!-\!\langle x \rangle \langle y \rangle$. Finally, $\langle \cdot \rangle$ denotes the average over all participants in one event. 
In Fig.~\ref{eps2} we show the eccentricity as a function of the collision's impact parameter. We show results for the time where corresponding to the two nuclei completely overlapping, estimated in a geometric way as:
\beq
t=R/(\sqrt{\gamma^2-1}),
\label{time}
\eeq
where $R$ is the nuclear radius and $\gamma$ is the Lorentz factor. 

We confirm that the impact of nucleon-nucleon correlations on $\varepsilon_2$ is negligible as demonstrated in~\cite{Alvioli:2011sk}, where correlations were shown to affect the fluctuations of flow harmonics. Further, $\varepsilon_2$ is shown to be resilient to the neutron skin effect (solid vs. dashed red lines in Fig.~\ref{eps2} and bottom pannel). This results from the fact that the neutron skin does not modify the global shape of the nucleus \textit{i.e.} the size of the nucleus remains identical with or without it. In turn, when focusing on the ratio of $\varepsilon_2$'s between the two isobar systems (Fig.~\ref{eps2}, bottom pannel) we observe up to a 10\% difference in ultra-central collisions. This effect persists down to mid-central collisions i.e. $b\!=\!4$~fm and it is not caused by neither the neutron skin nor the NN SRC. We pinpoint the deformation to be the source of this enhancement. This result at the eccentricity level is in quantitative agreement with the $v_2$ values shown in~\cite{Schenke:2019ruo}. Therefore, we suggest to only consider collisions with $b\!>\!6$~fm (to be translated into the experiment's centrality definition) in order to ensure an identical background component on the isobar run.

\subsection{Signal: Magnetic field strength}
Strong magnetic fields are essential to convert the chiral imbalance (see Eq.~\ref{Eq1}) into a discernible charge separation in the particles that reach the detector. Like previous works in the literature~\cite{Bzdak:2011yy,Skokov:2009qp} we compute the magnetic field in the framework of Lienard-Wiechert potentials~\cite{lienard,doi:10.1002/andp.19013090403},  \textit{i.e.}

\beq
e\vec{B}(t,\vec r)=\alpha\displaystyle\sum_{i=1}^{\rm N_{ch}}\displaystyle\frac{(1-v_i^2)(\vec v_i \times \vec R_i)}{R_i^3\left[1-(\vec R_i\times\vec v_i)^2 / R_i^2\right]^{3/2}},
\label{Bfield}
\eeq

\begin{figure}[ht]
\begin{center}
\includegraphics[scale=0.6]{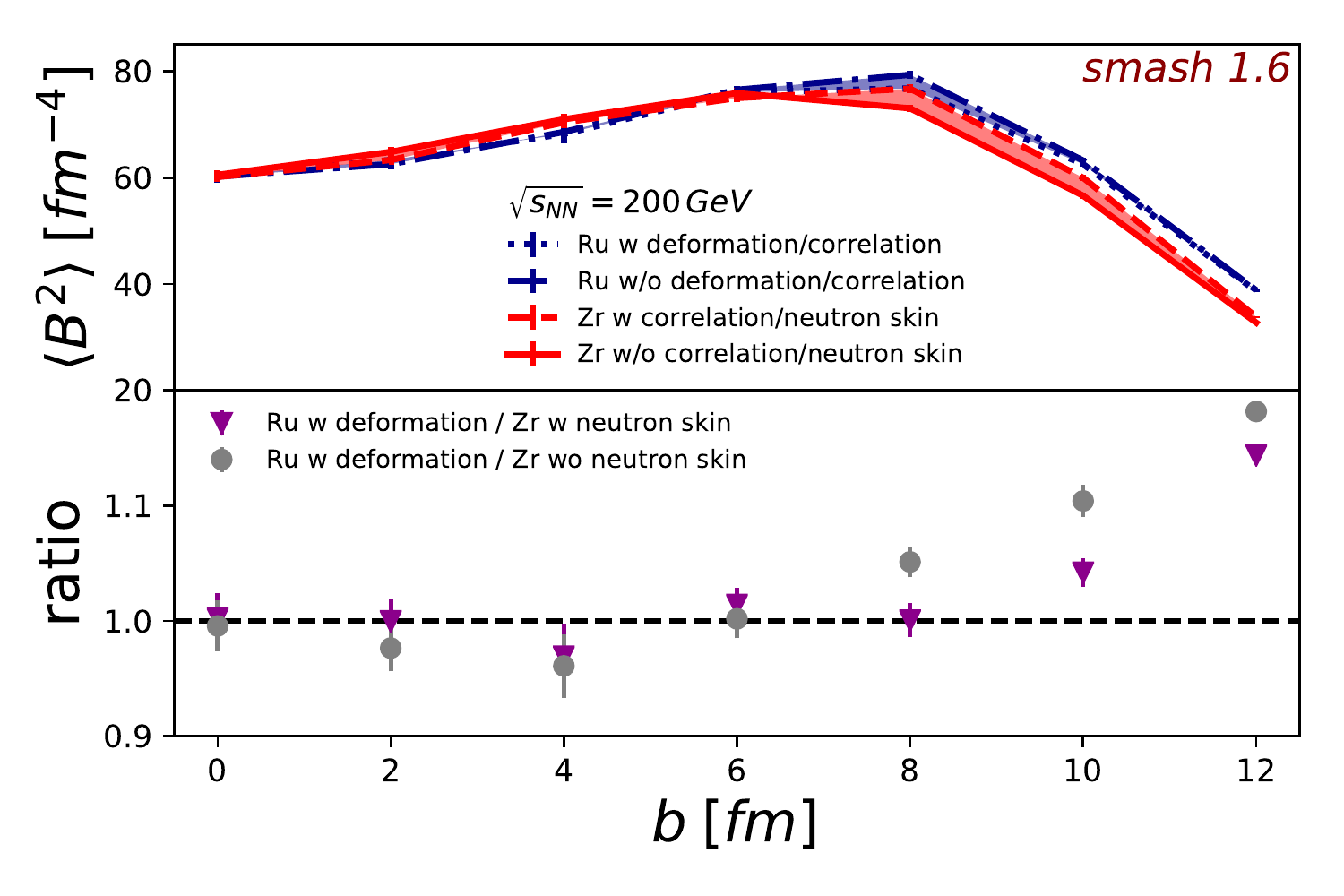} 
\end{center}
\vspace*{-0.5cm}
\caption[a]{Top: Strength of magnetic field (see Eq.~\ref{Bfield}) squared for $^{96}_{40}$Zr (red) with (solid) and without (dashed) neutron-skin and for $^{96}_{44}$Ru (blue) with (solid) and without (dashed) deformation. Bottom: Effect of the neutron skin on the magnetic field strength squared ratio between $^{96}_{44}$Ru and $^{96}_{40}$Zr.}
\label{Bsqrd}
\end{figure}

where the sum runs over all charged particles $\rm{N_{ch}}$, $\vec v$ is the velocity of each particle and $\vec R_i = \vec r-\vec r_i(t)$. In the last expression, $\vec r$ is the observation point and $\vec r_i$ the position of the $i$-th charged particle. We compute the magnetic field at the time where it is maximal, given by Eq.~\ref{time}, and at the central point $\vec r\!=\! 0$. To avoid singularities when $\vec R_i\rightarrow 0$, we do not include particles with $R_i \!<\!0.3$~fm in Eq.~\ref{Bfield}.

Figure~\ref{Bsqrd} shows the event-average magnetic field strength squared, $\langle B^2\rangle$, as a function of the impact parameter. Notice that, for completeness, in the top panel, we show the effect of deformation on the magnetic field in the $^{96}_{44}$Ru case. We refrain to compare this option with $^{96}_{40}$Zr as there are neither experimental indications nor theoretical predictions that suggest both nuclei to be undeformed. Therefore, in the bottom panel of Fig.~\ref{Bsqrd} we display the ratio between the two systems in a realistic scenario, i.e. considering deformation for $^{96}_{44}$Ru and focusing on the role of the neutron skin on the $^{96}_{40}$Zr case. We find that the inclusion of the neutron skin on the description of $^{96}_{40}$Zr's nuclear structure counterbalances the excess of protons in $^{96}_{44}$Ru and leads to a magnetic field strength ratio close to one in peripheral collisions ($12\!>\!b\!>\!8$~fm). We observe a sizeable difference on the magnetic field generated by both systems only arises when going to ultra-peripheral collisions ($b\!>\!12$~fm). This is the main result of this work that can be naturally interpreted as follows. The neutron skin, as shown in Fig.~\ref{1bratio}, enhances the number of neutron-neutron interactions in peripheral collisions or, equivalently, the concentration of protons in the central point that contribute to Eq.~\ref{Bfield}, leading to a larger $B$ field. Consequently, our study pushes the centrality cut needed to select the events where the CME search were to be performed to significantly larger values.
\section{Summary}
We investigated the influence of an experimentally measured feature of $^{96}_{40}$Zr, namely, the neutron skin effect on observables related to CME searches with the isobar program at RHIC. The main results of this work can be summarized as follows:
\begin{itemize}
    \item The background component, namely azimuthal correlations arising from flow, is expected to be $\mathcal O(10\%)$ larger in $^{96}_{44}$Ru+$^{96}_{44}$Ru than in $^{96}_{40}$Zr+$^{96}_{40}$Zr in ultra-central collisions.
    \item The difference between the magnetic field strength generated in both collision systems is reduced by half when including the neutron skin effect in the description of $^{96}_{40}$Zr. 
\end{itemize}
Therefore, we conclude that details of the nuclear spatial distributions need to be accounted for in a meaningful interpretation of the experimental measurements related to the CME effect in the isobar run.

\section*{ACKNOWLEDGMENTS} 
We would like to express our gratitude to Niklas Mueller, Vladimir Skokov and Pritwish Tribedy for helpful discussions during the realization of this work. M.A. acknowledges a CINECA award under ISCRA initiative for making high--performance computing resources available. H.E. and J.H.'s acknowledge support by the Helmholtz International Center for the Facility for Antiproton and Ion Research (HIC for FAIR) within the framework of the Landes-Offensive zur Entwicklung Wissenschaftlich- Oekonomischer Exzellenz (LOEWE) program from the State of Hesse. This project was further supported by the DAAD funded by BMBF with project-id 57314610 and by the Deutsche Forschungsgemeinschaft (DFG) through the grant CRC-TR 211 Strong-interaction matter under extreme conditions. A. S. O.'s work was supported by the U.S. Department of Energy, Office of Science, Office of Nuclear Physics, under contract No. DE- SC0012704, and by Laboratory Directed Research and Development (LDRD) funds from Brookhaven Science Associates. The research of M.S. was supported by the U.S. Department of Energy, Office of Science, Office of Nuclear Physics, under Award No. DE-FG02-93ER40771. Computing resources provided by the GOETHE-CSC are acknowledged as well. 

\appendix\section{Appendix A}
In order to transform the charge distribution to point like distributions of protons and neutrons we follow several steps based on~\cite{Patterson:2003wbr,Warda:2010qa,Salcedo:1987md}. First, we obtain the mean square charge distribution radius by using
\beq
\langle r^2_{\rm {ch}} \rangle = \displaystyle\frac{3R_0^2}{5} \Big(1+\displaystyle\frac{7\pi^2d^2}{3R_0^2}\Big).
\eeq
Next, the value of $\langle r^2_p \rangle$ is obtained by unfolding \textit{i.e.}
\beq
\langle r^2_{\rm {ch}} \rangle = \langle r^2_{\rm {p}} \rangle + R_p^2
\eeq
where the radius of the proton is $R_p\!=\!0.875$~\rm{fm}. After finding the value of $\langle r^2_{\rm {p}} \rangle$, one can calculate $R_{0,p}$ and $d_p$ as follows
\beq
R_{0,p}=R_0+\displaystyle\frac{5R_0\langle r_p^2\rangle}{7\pi^2d^2+15R_0^2},
\eeq
\beq
d_{p}^2=d^2-\displaystyle\frac{5\langle r_p^2\rangle(d^2+3R_0^2/\pi^2)}{7\pi^2d^2+15R_0^2}.
\label{difu}
\eeq
Once the Woods-Saxon parameters for the proton distribution are known, and in the case of a neutron-halo type ($R_{0,p}\!=\!R_{0,n}$), the only missing parameter is the diffusiveness of the neutron distribution. To find it, one has to replace $\langle r_p^2\rangle$ in Eq.~\ref{difu} by $(\Delta r_{np}+\langle r^2_p\rangle^{1/2})^2$.

This procedure leads to the values quoted in Table~\ref{ZrValues} that ensure the nucleus size to be identical when considering point-like or charge distributions.

\bibliography{references_neutronskin}{}
\bibliographystyle{apsrev4-1}

\end{document}